\documentclass{article} 
\usepackage{romp}
\usepackage[centertags]{amsmath}
\usepackage{color}
\usepackage{amsfonts}
\usepackage{amssymb}
\usepackage{amsthm}
\usepackage{newlfont}

\def\vveecc{\raise1.72ex\hbox{$\scriptscriptstyle\leftarrow$}\kern-.65em}
\begin{document}

\renewcommand{\theequation}{\arabic{section}\arabic{equation}}

\title{$\kappa$-Deformed Oscillators: Deformed Multiplication Versus Deformed Flip Operator and Multiparticle Clusters}

\author{
 J. Lukierski \\
Institute of Theoretical
 Physics, Wroc{\l}aw University, \\ 50-204 Wroc{\l}aw, Poland
 \\
 e-mail: lukier@ift.uni.wroc.pl}

\date{}

\maketitle

\begin{abstract}
We transform the oscillator algebra with $\kappa$-deformed
 multiplication rule, proposed in \cite{lukrep1,lukrep2}, into the  oscillator algebra with $\kappa$-deformed  flip operator and standard multiplication. We recall that the $\kappa$-multi\-pli\-ca\-tion of the $\kappa$-oscillators puts them off-shell. We study the explicit forms of modified mass-shell conditions in both formulations: with $\kappa$-multi\-pli\-ca\-tion and with $\kappa$-flip operation.
On the example of $\kappa$-deformed 2-particle states we study the clustered nonfactorizable form of the $\kappa$-deformed multiparticle states. We argue that the $\kappa$-deformed star product of two free fields leads in similar way to a nonfactorizable  $\kappa$-deformed bilocal field. We conclude with general remarks concerning the $\kappa$-deformed n-particle clusters and $\kappa$-deformed star product of $n$ fields.
\end{abstract}

\section{Introduction}

In this paper we shall investigate the class of $\kappa$-deformed oscillators and $\kappa$-deformed statistics \cite{lukrep1}--\cite{lukrep7a} which were introduced in  order to obtain the consistency  with $\kappa$-deformed addition law of the fourmomenta $p_\mu = (p_0,p_i)$. We shall consider the fourmomentum basis in $\kappa$-deformed Poincar\'{e} algebra (see e.g. \cite{lukrep7}-\cite{lukrep9}), with the following standard form of non-Abelian coproducts of the fourmomentum generators $P_\mu = (P_0, P_i)$.

 \def\theequation{\thesection1\alph{equation}}
 
\begin{eqnarray}\label{lukrep1.1a}
\Delta(P_i) & = &  P_i \otimes  e^{\frac{P_0}{2\kappa}}
+  e^{- \frac{P_0}{2\kappa}} \otimes P_i \, ,
\\
\Delta (P_0) & = &  P_0  \otimes  1  +  1 \otimes P_0 \,.\label{lukrep1.1b}
\end{eqnarray}

We introduce the $\kappa$-deformed bosonic oscillators 
$a_\kappa (p) \equiv a_\kappa( p_0, {\vec{p}})$ satisfying the relation ($p_\mu = p_0, \vec{p}$)
 \def\theequation{\thesection\arabic{equation}}
 \setcounter{equation}{1}
\begin{equation}\label{lukrep1.2}
P_\mu \triangleright a_\kappa(p) \equiv adj_{P_\mu} a_\kappa (p)
= p_\mu a_\kappa (p) \,,
\end{equation}
with 
 the quantum adjoint, generalizing the commutator,  given by the formula
$ad\, j_{A} B = A_{(1)} BS(A_{(2)})$,  where $\Delta(A)=A_{(1)} \otimes A_{(2)}$. We apply  the following Hopf algebraic action formula

\begin{eqnarray}\label{lukrep1.3}
P_\mu \triangleright (a_\kappa (p) \, a_\kappa(q)) 
& = & \left( P_{\mu (1)}  \triangleright  a_\kappa(p)\right) \cdot
\left(P_{\mu (2)} \triangleright  a_\kappa (q)\right)
\cr
 &= &
p^{(1+2)}_\mu  a_\kappa (p) a_\kappa(q) \,,
\end{eqnarray}
where from (\ref{lukrep1.1a},b)-(\ref{lukrep1.2}) one gets

 \def\theequation{\thesection4\alph{equation}}
  \setcounter{equation}{0}
\begin{eqnarray}\label{lukrep1.4a}
p^{(1+2)}_i & = &  p_i\, e^{\frac{q_0}{2\kappa}} + q_i \, e^{-\frac{p_0}{2\kappa}}
\\
p_0^{(1+2)} & = & p_0 + q_0  \, .
\label{lukrep1.4b}
\end{eqnarray}
Exchanging in (\ref{lukrep1.3}) the values of $p_\mu$ and $q_\mu$ we obtain the fourmomentum eigenvalues for flipped 2-particle state

\def\theequation{\thesection5\alph{equation}}
  \setcounter{equation}{0}
\begin{eqnarray}\label{lukrep1.5a}
p_i^{(2+1)} & = & q_i \, e^{\frac{p_0}{2\kappa}}
+ p_i \, e^{-\frac{q_0}{2\kappa}}
\\
p^{(2+1)}_0 & = & q_0 + p_0 = p_0 ^{(1+2)} \, .
\label{lukrep1.5b}
\end{eqnarray}
From the comparison of relations (\ref{lukrep1.4a}) and
(\ref{lukrep1.5a}) it follows easily that  for $\kappa < \infty$  we get
$p_i^{(1+2)} \neq p_i^{(2+1)}$.  We see therefore that the standard exchange relation for the creation operators $a_\kappa(\vec{p}, p_0) \equiv a_\kappa(p)$

\def\theequation{\thesection\arabic{equation}}
  \setcounter{equation}{5}
\begin{equation}\label{lukrep1.6}
a_\kappa(p) \, a_\kappa(q) = a_\kappa(q)\, a_\kappa(p)
\end{equation}
is inconsistent with the three-momentum addition law (\ref{lukrep1.1a}).

It appears that the relation (\ref{lukrep1.6}), as well as remaining standard commutation relations for the $\kappa$-deformed creation and annihilation operators $a_\kappa(p_0, \vec{p}),a^+_\kappa(p_0,\vec{p})$ should be modified. First
proposal which is consistent with 
 the $\kappa$-deformed fourmomentum addition law 
  (\ref{lukrep1.3}--\ref{lukrep1.4a},b) has been presented in \cite{lukrep1}, and it is described by the following set of relations:

\begin{equation}\label{lukrep1.7}
a_\kappa \left({ e}^{-\frac{q_0}{2\kappa}}\vec{p}, p_0 \right)
a_\kappa \left({ e}^{\frac{p_0}{2\kappa}}\vec{q},q_0 \right) =
a_\kappa \left({ e}^{-\frac{p_0}{2\kappa}}\vec{q}, q_0 \right)
a_\kappa \left({ e}^{\frac{q_0}{2\kappa}}\vec{p}, p_0\right)
\;,
\end{equation}

\begin{equation}\label{lukrep1.8}
a_\kappa^\dag \left({ e}^{\frac{q_0}{2\kappa}}\vec{p}, p_0 \right)
a_\kappa^\dag \left({ e}^{-\frac{p_0}{2\kappa}}\vec{q}, q_0 \right)
 = 
a_\kappa^\dag \left({e}^{\frac{p_0}{2\kappa}}\vec{q},q_0 \right) 
a_\kappa^\dag \left({e}^{-\frac{q_0}{2\kappa}}\vec{p}, p_0\right) \; . 
\end{equation}

\begin{equation}\label{lukrep1.9}
a_\kappa^\dag \left({ e}^{-\frac{q_0}{2\kappa}}\vec{p}, p_0\right) 
a_\kappa \left({ e}^{-\frac{p_0}{2\kappa}}\vec{q}, q_0\right) 
- 
a_\kappa \left( {e}^{\frac{p_0}{2\kappa}}\vec{q}, q_0 \right) 
a_\kappa^\dag \left({e}^{\frac{q_0}{2\kappa}}\vec{p}, p_0\right) 
= \delta^{(3)}\left(\vec{p} -
\vec{q} \right)\;.
\end{equation}
where
\begin{equation}\label{lukrep1.10}
p_0 = \omega_\kappa (\vec{p}) \qquad q_0= \omega_\kappa(\vec{q})\, ,
\end{equation}
and $\omega_\kappa$ is the solution of $\kappa$-deformed mass-shell condition
\begin{equation}
\vec{p}^{\ 2}-\left(2\kappa~ {\textrm sinh}\frac{p_0}{2\kappa}\right)^2=m_0^2\, .
\end{equation}
Explicitely we obtain
\begin{equation}
\omega_\kappa(\vec{p})=2\kappa~ {\textrm Arcsinh}\left[\frac{(\vec{p}^{\ 2}+m_0^2)^{1/2}}{2\kappa}\right]\, .
\end{equation}
The three-momenta describing the oscillators in the relations (\ref{lukrep1.7}-\ref{lukrep1.9}) are put therefore in particular way off-shell.

The $\kappa$-deformed oscillator algebra (\ref{lukrep1.7}-\ref{lukrep1.9}) 
can
be rewritten in standard algebraic form if we define the following
$\kappa$-multiplication of the oscillators $a _\kappa(p )$
\begin{equation}\label{lukrep1.11}
a_\kappa (p)\circ a_\kappa(q):=a_\kappa \left({
e}^{-\frac{q_0}{2\kappa}}\vec{p}, p_0 \right) 
a_\kappa \left({e}^{\frac{p_0}{2\kappa}}\vec{q}, q_0 \right)\;, 
\end{equation}
and if, in consistency with the conjugation relation
$a_{\kappa}^\dag (p) = a_{\kappa}(-p)$, we introduce
\begin{equation}\label{lukrep1.12}
a_\kappa^\dag (p)\circ a_\kappa^\dag (q)
= 
a_\kappa^\dag\left({ e}^{\frac{q_0}{2\kappa}}\vec{p}, p_0 \right) 
a_\kappa^\dag
\left({ e}^{-\frac{p_0}{2\kappa}}\vec{q}, q_0 \right)\;,\;\;\;\;\;
\end{equation}

\begin{equation}\label{lukrep1.13}
a_\kappa^\dag (p)\circ a_\kappa(q)= a_\kappa^\dag \left({
e}^{-\frac{q_0}{2\kappa}}\vec{p}, p_0 \right) 
a_\kappa \left({
e}^{-\frac{p_0}{2\kappa}}\vec{q}, q_0 \right)\;,\;\; 
\end{equation}
\vspace{0.03cm}
\begin{equation}\label{lukrep1.14}
a_\kappa (p)\circ a_\kappa^\dag(q)= 
a_\kappa\left({
e}^{\frac{q_0}{2\kappa}}\vec{p}, p_0, \right) 
a_\kappa^\dag \left({
e}^{\frac{p_0}{2\kappa}}\vec{q}, q_0 \right)\;.
\;\;\;\;\;\;
\end{equation}

We see that the relations (\ref{lukrep1.7})-(\ref{lukrep1.9}) 
 can be described in the form of 
 standard algebra of creation and annihilation operators
$([\;A,B\;]_\circ := A\circ B - B\circ A)$
\begin{equation}\label{lukrep1.15}
[\;a_\kappa  (p),a_\kappa  (q)\;]_{\circ} =  [\;a_\kappa ^\dag
(p),a_\kappa ^\dag (q)\;]_{\circ} = 0\;\;,\;\;[\;a_\kappa ^\dag
(p),a_\kappa  (q)\;]_{\circ} = \delta^{(3)} (\vec{p}-\vec{q})\;.
\end{equation}
The effect of the $\kappa$-deformation on the algebra
 ${\cal A}(a,a^+)$  of classical creation and annihilation operators consists in the replacement of standard binary multiplication rule by the 
 $\kappa$-deformed one
 \begin{equation}\label{lukrep1.16}
 m(A \otimes B) = A \cdot B 
 \mathop{\longrightarrow}\limits_{ \kappa < \infty }
 m_\kappa (A\otimes B) = A \circ B\,.
 \end{equation}

It appears however that one can introduce the $\kappa$-deformation of the oscillator algebra in an alternative way by introducing the flip operator which modifies the notion of commutator. In such an approach the
 standard multiplication operation remains unchanged. The standard commutator  $[A,B]\equiv A \cdot B - B \cdot A$ can be described as follows
\begin{equation}\label{lukrep1.17}
	[A,B] = (1 - \hat{\tau}_0 )A\cdot B\,,
\end{equation}
where the standard flip operator $\hat{\tau}_0 $ reverses the order of the elements in the product
\begin{equation}\label{lukrep1.18}
	{\tau}_0 (A\cdot B)  = B \cdot A \qquad \hat{\tau}_0^2 = 1\,. 
\end{equation}
One can define the $\kappa$-deformed oscillator algebra by introducing  the $\kappa$-de\-for\-ma\-tion of the flip operator $\hat{\tau}_0 $, i.e. we replace (see e.g. \cite{lukrep4}--\cite{lukrep7a},\cite{lukrep10})

\begin{equation}\label{lukrep1.19}
	[A, B] \mathop{\longrightarrow}\limits_{ \kappa < \infty }
	[A, B]_\kappa = (1 - \hat{\tau}_\kappa) A\cdot B 
\end{equation}
where
\begin{equation}\label{lukrep1.20}
	\hat{\tau}^2_\kappa = 1 \, .
\end{equation}
The correct choice, consistent with the addition law (\ref{lukrep1.4a}-b) leads to the condition
\begin{equation}\label{lukrep1.21}
	[\hat{\tau}_\kappa, \Delta(P_\mu)]=0\,.
\end{equation}
We modify the standard bosonic exchange relation (\ref{lukrep1.6}) as follows
\begin{equation}\label{lukrep1.22}
	a_\kappa(p) a_\kappa(q) =
	\hat{\tau}_\kappa 
	(a_\kappa(p) a_\kappa(q))\,.
\end{equation}
Applying (\ref{lukrep1.3}) to both sides of (\ref{lukrep1.22}) and using (\ref{lukrep1.21}) we obtain that the action of the fourmomentum operator $P_\mu$ provides the same value of two-particle fourmomenta.

First aim of this paper is to show (see Sect.~2) that our $\kappa$-deformed oscillator algebra presented in \cite{lukrep1,lukrep2} can be rewritten in the form (\ref{lukrep1.22}), with suitable extension of $\kappa$-deformed flip to remaining binary oscillator products. For this purpose we shall use the nonlinear transformations in two-particle fourmomenta space, which changes as well the explicit form of the fourmomentum addition law.
 In such a way modulo modification of the $\kappa$-Poincar\'{e} algebra basis we obtain the form of $\kappa$-deformed  oscillator algebra, which recalls  the results of Arzano and Marciano \cite{lukrep3}, however with different energy-momentum dispersion relations.
 We stress that
 due to proposed $\kappa$-deformed energy-momentum dispersion relations, in two-particle state 
 the energies $p_0, q_0$ of each of the particles depends on the three-momenta of both particles. The additional dependence on second three-momentum is generated by the $\kappa$-deformation, which introduces some geometric two-particle interactions. 
 
 In Sect.~3 just on the example of $\kappa$-deformed two-particle states we shall show that the notion of $\kappa$-multiplication destroys the factorizability into single particle excitations,  i.e. two-particle $\kappa$-deformed states describe inseparable clustered states. We shall show that such   nonfactorizable \, $\circ$-multiplication of states is related with the property of a $\star_\kappa$-multiplication of free fields. Because one can treat the\,  $\circ$-multiplication of the oscillators in momentum space as dual to the $\star_\kappa$-product multiplication of fields in space-time, we shall point out as well the nonfactorizability of the star product of quantum fields.

Finally in Sect.~4 we recall the extension of the definition of  
  \, $\circ$-multi\-pli\-cation to n-fold products of oscillators consistent with the associativity property. We discuss briefly the unusual features of such an multiplication, which leads for various $n$ to different forms of energy-momentum relations in the oscillator binary relations
   if they are  factorized from the products of $n$ ($n>2$) oscillators.
   The statistics involving $n$ oscillators ($n>2$) we discuss in Sect.~4 only in the framework using $\kappa$-multiplication. In general case 
    the product of $n$ oscillators 
   in the formalism using $\kappa$-deformed flip operator 
     requires  modification of the assumption that the arbitrary  $n$-particle flip operator decomposes into the product of two-particle flip operators representing the  geometric 2-body forces  (see e.g. \cite{lukrep6,lukrep7a}). Further discussion of these problems we shall present in near future.

\section{The $\kappa$-deformed Oscillator Algebra with $\kappa$-deformed Flip Operator}

\setcounter{equation}{0}

The  $\kappa$-deformed algebra (\ref{lukrep1.15}) has been arranged in such a way that the two-particle states  defined as follows ($p_0=\omega_\kappa(\vec{p}); \, q_0 = \omega_\kappa(\vec{q}))$ \, \cite{lukrep1}

\begin{equation}\label{lukrep2.1}
\| \vec{p}, \vec{q}>\!\!>_\kappa =
 a_\kappa (\vec{p}, p_0)
 \circ a_\kappa(\vec{q}, q_0)|0\rangle 
\end{equation}
where
\begin{equation}\label{lukrep2.2x}
	a^+_\kappa(\vec{p}, p_0)|0\rangle = 0
\end{equation}
 carry the classical two-momentum value:
\begin{eqnarray}\label{lukrep2.2}
	P_\mu \| \vec{p}, \vec{q}>\!\!>_\kappa & = &
	P_\mu \triangleright (a_\kappa(\vec{p},p_0) \circ a_\kappa
	(\vec{q}, q_0))|0\rangle
	\cr
	& = & P_\mu \triangleright (a_\kappa(\vec{p} \, e^{-\frac{q_0}{2\kappa}}, p_0)
	a_\kappa(\vec{q}\,  e^{\frac{p_0}{2\kappa}}, q_0)|0\rangle
	\cr
	& = & (p_\mu + q_\mu) \|\vec{p}, \vec{q} \, \rangle_\kappa \,.
\end{eqnarray}

One can say that in the state (\ref{lukrep2.1}) the  three-momentum dependence of the creation oscillators $a(\vec{p}, p_0)$ is modified exactly in a way which compensates the  deformation of classical Abelian addition law described by  (\ref{lukrep1.4a}). 
We see that however we use the $\kappa$-deformed non-Abelian addition law (\ref{lukrep1.4a}),  the  fourmomenta  of two $\kappa$-deformed particles add in Abelian way (see (\ref{lukrep2.2})).
Due to the relation (see (\ref{lukrep1.15}))
\begin{equation}\label{lukrep2.4x}
	a(p) \circ a(q) = a(q) \circ a(p)
\end{equation}
the states (\ref{lukrep2.1}) are endowed with the standard bosonic symmetry   properties
\begin{equation}\label{lukrep2.3}
	\| \vec{p}, \vec{q}>\!\!>_\kappa = \| \vec{q}, \vec{p}>\!\!>_\kappa \,.
\end{equation}

In order to see in different way the effect of the exchange of two  $\kappa$-deformed particles with fourmomenta $p_\mu, q_\mu$ one can consider as well the form (\ref{lukrep1.19}) of the $\kappa$-deformed oscillator algebra. In such a case the $\kappa$-deformed flip operator should change the values $\vec{p}, \vec{q}$ in flipped 2-particle state. We introduce the two-particle states as follows:
\begin{eqnarray}\label{lukrep2.4}
	|\vec{p}\, ' , \vec{q}\, '\rangle{_\kappa} & = &
	a_\kappa(\vec{p}\, ', \omega _\kappa(\vec{p}\, '))
	a_\kappa (\vec{p}\, ', \omega_\kappa(\vec{q}\, '))|0\rangle
	\\
		|\vec{q}\, ' , \vec{p}\, '\rangle_{\tau_\kappa} & = &
	\tau_\kappa [a_\kappa(\vec{p}\, ', \omega _\kappa(\vec{p}))
	a_\kappa (\vec{q}, \omega_\kappa(\vec{p}\, '))]|0\rangle \,.
	\label{lukrep2.5x}
\end{eqnarray}
The algebra of $\kappa$-deformed oscillators (\ref{lukrep1.22}) leads to the identification of the states
\begin{equation}\label{lukrep2.5}
		|\vec{p}\, ' , \vec{q}\, '\rangle_\kappa = 	|\vec{q}\, ' , \vec{p}\, ' \rangle_{\tau_\kappa} \, . 
\end{equation}
Introducing the symmetric states

\begin{equation}\label{lukrep2.6}
		|\vec{p}\, ' , \vec{q}\, ' \rangle_s = 
		\frac{1}{2}
			\left(
			|\vec{p}\, ' , \vec{q}\, ' \rangle_\kappa
			+ | \vec{p}\, ' , \vec{q} \, ' \rangle_{\tau_\kappa} \right)
\end{equation}
due to (\ref{lukrep1.20}) we obtain the form of the relation (\ref{lukrep2.5}) analogous to (\ref{lukrep2.3})

\begin{equation}\label{lukrep2.7}
		|\vec{p}\, ' , \vec{q}\, ' \rangle_s = 	|\vec{q}\, ' , \vec{p}\, ' \rangle_{s} \, . 
\end{equation}

In order to relate the states (\ref{lukrep2.1}) and (\ref{lukrep2.4}) we should introduce the following transformation of the pair of three-momentum coordinates
\begin{equation}\label{lukrep2.8}
	\vec{p}\, ' = \vec{p}\, e^{- \frac{q_0}{2\kappa}} \qquad
	\vec{q}\, ' = \vec{q}\, e^{\frac{p_0}{2\kappa}} \, ,
\end{equation}
and insert these new values $\vec{p}\, ' , \vec{q}\, '$ into the $\kappa$-deformed energy-momentum dispersion relation (\ref{lukrep1.10}).
 The dispersion relations (\ref{lukrep1.10})
in new variables $\vec{p}\, ', \vec{q}\, '$ describe the following pair of coupled $\kappa$-deformed mass-shell conditions describing the energy values $p_0\, ',q_0\, '$
\begin{eqnarray}\label{lukrep2.9}
p'_0  & = & p_0 = \omega_ \kappa(\vec{p})=
\omega_\kappa (\vec{p}\, ' \, e^{\frac{q'_0}{2\kappa}})
\equiv \omega^{(+)}_\kappa(\vec{p}\, ')
\cr
q'_0  & = & q_0 = \omega_ \kappa(\vec{q})=
\omega_\kappa (\vec{q}\, ' \, e^{-\frac{p\, '_0}{2\kappa}})
\equiv \omega^{(-)}_\kappa(\vec{q}\, ')
\end{eqnarray}
where $(\varepsilon = \pm 1, \eta=\pm1)$
\begin{equation}\label{lukrep2.10}
	\omega^{(\varepsilon)}_\kappa(\vec{p}) = \omega_\kappa(\vec{p}\, e^{\varepsilon \frac{q_0}{2\kappa}})
	\qquad
		\omega^{(\eta)}_\kappa(\vec{q}) = \omega_\kappa(\vec{q}\, e^{\eta \frac{p_0}{2\kappa}}) \, .
\end{equation}

After substituting (\ref{lukrep2.8}) it is easy to see from (\ref{lukrep1.7}) that one gets the relations (\ref{lukrep1.22}) with the following explicit form of the $\kappa$-deformed flip operator
\begin{equation}\label{lukrep2.10a}
	\tau_\kappa
	\left(
	a_\kappa(\vec{p}\, ', p'_0) a_\kappa(\vec{q}\, ' , q'_0)\right)
	=
	 a_\kappa\left(\vec{q}\, ' \, e^{- \frac{p '_0}{\kappa}} \, , q'_0 \right)
	a_\kappa
	\left(
	\vec{p}\, ' \, e^{\frac{q '_0}{\kappa}}, p'_0
	\right)\, .
\end{equation}
One can check that both sides of the relation (\ref{lukrep1.22}) provide the same 2-particle non-Abelian sum of three-momenta

\begin{equation}\label{lukrep2.11}
	\vec{p}_{(1+2)} = \vec{p}\, ' \, e^{\frac{q'_0}{2\kappa}}
	+ 
	\vec{q}\, ' \, e^{- \frac{p'_0}{2\kappa}} = \vec{p}_{(2+1)}
\end{equation}
as well as the same two particle energies
\begin{equation}\label{lukrep2.12}
	p_0^{(1+2)} = \omega^{(-)}_\kappa (\vec{p}\, ') +
	\omega^{(+)}_\kappa (\vec{q}\, ') = p_0^{(2+1)}
\end{equation}
where $p'_0, q'_0$ are the solutions of algebraic relations (\ref{lukrep2.9}) provided by two energy functions $p'_0 = p'_0(\vec{p} \, ' , \vec{q} \, ')$ and
$q'_0 = q'_0(\vec{p}\, ', \vec{q} \, ')$.

The flip operator $\tau_\kappa$ can be also derived for other binary 
 $\kappa$-deformed products $a a^+, a^+ a$ and $a^+ a^+$ 
 (see (\ref{lukrep1.12})--(\ref{lukrep1.14})) 
  and one can as well rewrite all the $\kappa$-deformed relations (\ref{lukrep1.15}) in flipped form.
 Inserting in(\ref{lukrep1.8})

\begin{equation}\label{lukrep2.13}
	\vec{p}\, ' = \vec{p} \, e^{\frac{q_0}{2\kappa}} \qquad
	\quad
	\vec{q}\, ' = \vec{q} \, e^{- \frac{p_0}{2\kappa}}
\end{equation}
one gets

\begin{equation}\label{lukrep2.14}
	a^+_\kappa (\vec{p} \, ' , {p} \, '_0)\cdot
	a^+_\kappa(\vec{q}\,' , q\, '_0)
	=
	a^+_\kappa (\vec{q} \, ' \,
	e^{\frac{p '_0}{\kappa}}
	 , {q}\, '_0)
	 a^+_\kappa (\vec{p} \, ' \,
	e^{- \frac{q '_0}{\kappa}}
	 , {p}\, '_0) 
\end{equation}
where
\begin{equation}\label{lukrep2.15}
	p_0\, ' = \omega^{(-)}_\kappa (\vec{p}\, ')
	\qquad
	q_0\, ' = \omega^{(+)}_\kappa (\vec{q}\, ')\, .
\end{equation}
Further by putting
\begin{equation}\label{lukrep2.16}
	\vec{p}\, ' = \vec{p} \, e^{-\frac{q_0}{2\kappa}}
	\qquad
		\vec{q}\, ' = \vec{q} \, e^{-\frac{p_0}{2\kappa}}
\end{equation}
one gets
\begin{equation}\label{lukrep2.17}
	a^+_\kappa (\vec{p} \, ' , {p} \, '_0)
	a_\kappa (\vec{q} \, ' 
	 , {q} \, '_0)
	 -
	 a_\kappa (\vec{q} \, ' \,
	e^{\frac{p\, '_0}{\kappa}}
	 , {q} \, '_0) 
	  a^+_\kappa (\vec{p} \, ' \,
	e^{\frac{q\, '_0}{\kappa}}
	 , {p}\, '_0) =
	 \delta^3
	  (\vec{p} \, ' \,
	e^{\frac{q\, '_0}{2\kappa}}
	- \vec{q}\, '_0\,  e^{\frac{p\, '_0}{2\kappa}})
\end{equation}
where in this case $p'_0, q'_0$ satisfies two coupled equations
\begin{equation}\label{lukrep2.18}
	p\,'_0 = \omega^{(+)}_\kappa(\vec{p}\,')
	\qquad
		q\,'_0 = \omega^{(+)}_\kappa(\vec{q}\,') \, .
\end{equation}
The $\kappa$-deformed flip operator acts therefore on   arbitrary  binary products of $\kappa$-deformed oscillators as follows
\begin{equation}
\begin{array}{l}
\label{lukrep2.19}
  {\small \hat{\tau}_\kappa
	\begin{pmatrix}
a_\kappa(\vec{p},p_0)a_\kappa(\vec{q},q_0)
	&
	a_\kappa(\vec{p}, p_0)a^+_\kappa(\vec{q}, q_0)
	\cr
		a^+_\kappa(\vec{p}, p_0)a_\kappa(\vec{q}, q_0)
	&
	a^+_\kappa(\vec{p}, p_0)a^+_\kappa(\vec{q}, q_0) 
	\end{pmatrix} = }
\cr\cr
{\small 
	 =\begin{pmatrix}
	a_\kappa(\vec{q} e^{-\frac{p_0}{\kappa}},\omega^{(-)} (\vec{q}))
	a_\kappa(\vec{p} e^{\frac{q_0}{\kappa}}, \omega^{(+)}(\vec{p}))
	& 
	a^{+}_\kappa(\vec{q} e^{-\frac{p_0}{\kappa}},
	\omega^{(-)} (\vec{q}))
	a(\vec{p} e^{-\frac{q_0}{\kappa}}, \omega^{(-)}(\vec{p}))
	\cr
	a_\kappa(\vec{q} e^{\frac{p_0}{\kappa}},
	\omega^{(+)}_\kappa (\vec{q}))
	a^{+}_\kappa(\vec{p} e^{\frac{q_0}{\kappa}}, \omega^{(+)}(\vec{p}))
		&
		a^{+}_\kappa(\vec{q} e^{\frac{p_0}{\kappa}},
	\omega^{(+)} (\vec{q}))
	a^{+}(\vec{p} e^{-\frac{q_0}{\kappa}}, \omega^{(-)}(\vec{p}))
	\end{pmatrix}
	}
\end{array}
\end{equation}
It can be checked that for the choice (\ref{lukrep2.19}) the relation (\ref{lukrep1.20}) is valid. 

From the relations  (\ref{lukrep1.22}), (\ref{lukrep2.14}) and (\ref{lukrep2.17}) we see that the simple binary products $a_\kappa(\vec{p}, p_0)a_\kappa(\vec{q}, q_0)$,
$a^+_\kappa(\vec{p}, p_0)a(\vec{q}, q_0)$ and
$a^+(\vec{p}, p_0)a^+(\vec{q}, q_0)$ are cha\-rac\-terized by different energy-momentum relations. Introducing the notation

\begin{equation}\label{lukrep2.20a}
	a(\vec{p}, p_0) = a^{(1)}(\vec{p}, p_0),\qquad
	a^+(\vec{p}, p_0) = a^{(-1)}(\vec{p}, p_0)
\end{equation}
one gets the following choice of the   mass-shell conditions in arbitrary binary products ($\varepsilon=\pm1, \eta=\pm1$, see (\ref{lukrep2.10}))
\begin{equation}\label{lukrep2.20}
	a^{(\varepsilon)} (\vec{p}, \omega^{(\eta)}_\kappa(\vec{p}))
	a^{(\eta)} (\vec{q}, \omega^{(-\varepsilon)}_\kappa(\vec{q}))
\end{equation}
in accordance with the formulae (\ref{lukrep2.10}), (\ref{lukrep2.15}) and (\ref{lukrep2.18}). In particular the relation (\ref{lukrep2.17}) can be written equivalently as follows
\begin{eqnarray}\label{lukrep2.21}
a_\kappa(\vec{p}\, ', \omega^{(+)}_\kappa(\vec{p}\, '))
a^+(\vec{q}\, ', \omega^{(-)}_\kappa(\vec{q}\, '))
-
a^+_\kappa(  \vec{q}\, '\,e^{-\frac{p\, '_0}{\kappa}}
, \omega^{(-)}_\kappa (\vec{q})) &&
\cr
\cdot
a_\kappa(  \vec{p}\, '\,e^{\frac{q '_0}{\kappa}}
, \omega^{(+)}_\kappa (\vec{p}\, '))
=
\delta^3 (\vec{p}\, ' e^{-\frac{q '_0}{2\kappa}}
- \vec{q}\, ' 
e^{\frac{p '_0}{2\kappa}}
)
&&
\end{eqnarray}
and by nonlinear transformation
\begin{equation}\label{lukrep2.22}
	\vec{p}\, '' = 
	e^{\frac{q '_0}{\kappa}} \vec{p}\, ' \qquad \qquad
		\vec{q}\, '' = 
	e^{-\frac{p\, ' _0}{\kappa}} \vec{q}\, '
\end{equation}
can be transformed into the relation (\ref{lukrep2.17}).

We see therefore that all four relations (\ref{lukrep1.22}), (\ref{lukrep2.14}), (\ref{lukrep2.17}) and (\ref{lukrep2.21})
 are written in terms of the $\kappa$-deformed oscillators carrying  energy described by different $\kappa$-deformed energy-momentum relations. The choice of the particular dispersion relation (\ref{lukrep2.10}) depends on the choice of the  partner in the bilinear product (see (\ref{lukrep2.20})). Different choices of the energy-momentum relations is the effect due to the $\kappa$-deformation and these choices differ by the sign of leading $\frac{1}{\kappa}$ terms. It should be added that these modified dispersion relations 
  are adjusted in such a way that they 
 provide  the conservation of energy (see e.g. (\ref{lukrep2.12})) under the flip operation. It can be recalled that for the product of $\kappa$-deformed creation operators Arzano and Marciano \cite{lukrep3} proposed in their construction of $\kappa$-deformed 2-particle states  the relations similar to (\ref{lukrep1.22}), however with the energies of all oscillators satisfying the $\kappa$-deformed standard mass shells (\ref{lukrep1.10}). Such a framework is described by the on-shell oscillators  but 
  unfortunately it 
 leads to the $\kappa$-statistics which does not preserve the two-particle energy under the exchange (flip operation) of two constituents of $\kappa$-deformed 2-particle states.

\section{Nonfactorizable Structure of $\kappa$-deformed Two-particle Clusters}

\setcounter{equation}{0}

Le us recall that the standard bosonic Fock space is a sum of symmetrized tensor products of one-particle states:

\begin{equation}\label{lukrep3.1}
	{\cal F} = \sum\limits^{\infty}_{n=0}
	{\cal H}^{(0)}_n
	\qquad  \quad
	{\cal H}^{(0)}_n = S_n({\cal H}^{(0)}_1 \otimes
	\ldots \otimes {\cal H}^{(0)}_1)
\end{equation}
where the one-particle states are generated from the vacuum ${\cal H}^{(0)}_0\equiv |0\rangle$ by the standard creation  operators. Subsequently, the states describing the basis of ${\cal H}^{(0)}_n$  
 are  generated by the product of $n$ creation oscillators
\begin{equation}\label{lukrep3.2}
	|p_1 \ldots p_n \rangle = a (p_1) \ldots a(p_n) |0\rangle\, .
\end{equation}
The standard tensor structure of ${\cal H}^{(0)}_n$ is ensured by the relation (\ref{lukrep1.6}). Concluding, in ${\cal H}^{(0)}_n$ 
 one-particle states in 
 every factor ${\cal H}^{(0)}_1$ (see (\ref{lukrep3.1})) can be changed independently from  the parameters of remaining $n-1$ one-particle  states.

In the case of $\kappa$-deformed oscillators the relation (\ref{lukrep1.6}) is replaced by the $\kappa$-deformed one - given by (\ref{lukrep1.7}) or by (\ref{lukrep1.22}), with $\tau_\kappa$ described by (\ref{lukrep2.10}). We shall consider further for simplicity the 2-particle states.

In $\kappa$-deformed Fock space
\begin{equation}\label{lukrep3.3}
{\cal F}^\kappa = \sum\limits^{\infty}_{i=0} {\cal H}^\kappa_n
	\end{equation}
the general formula for  two-particle sector takes the form
\begin{equation}\label{lukrep3.4}
{\cal H}^\kappa_2 = S^\kappa_2 ({\cal H}_1 \otimes_\kappa {\cal H}_1)\, .
	\end{equation}
	
The two choices of $\kappa$-deformed oscillators given by (\ref{lukrep1.11})
 and (\ref{lukrep2.17}) correspond to the following special choices
 
 \begin{description}
 \item{1)} $\kappa$-deformed multiplication (see (\ref{lukrep1.11}))
 
\begin{equation}\label{lukrep3.5}
S^\kappa_2 = S_2  \,.
	\end{equation}	
	We define new binary tensor product $\otimes_\kappa$ of one-particle states by the formula (\ref{lukrep2.1})
	\begin{equation}
	|\vec{p}>\otimes_\kappa|\vec{q}>\equiv a_\kappa(\vec{p}\,,p_0)\circ a_\kappa(\vec{q}\,,q_0)|0>\,.\label{lukrep3.5a}
	\end{equation}
	The introduction of $\kappa$-deformed 
 multiplication in (\ref{lukrep3.5a}), contrary to the case of standard Fock space, does not permit the factorization into one-particle states of the two-particle wave packet. It also follows from the relation (\ref{lukrep2.3}) and relation (\ref{lukrep3.5}).
	
	\item{2)} $\kappa$-deformed flip operator (see (\ref{lukrep1.22}), (\ref{lukrep2.14}) and (\ref{lukrep2.17}))
	\begin{equation}\label{lukrep3.6}
	S^\kappa_2 = \frac{1}{2} (1 \otimes 1 + \hat{\tau}_\kappa)
	\qquad \qquad \otimes_\kappa = \otimes
\end{equation}
\end{description}
i.e. we introduce the $\kappa$-deformation through the modification of the symmetrization procedure.

From the relation (\ref{lukrep1.7})  follows that the products  of one-particle wave  packet creators
\begin{equation}\label{lukrep3.7}
	\left(\int d^3\vec{p}\, f^{(1)} (\vec{p}) a(\vec{p}) \right)
	\circ_\kappa
	\left(\int d^3\vec{q}\, g^{(1)} (\vec{q}) a(\vec{q}) 
	\right)
	\end{equation}
	 can not be well defined, because the $\kappa$-multiplication in (\ref{lukrep3.7}) does not commute with the momenta integrations.
		If we introduce $f^{(2)}(p,q)=f^{(1)}(\vec{p})\, f^{(1)}(\vec{q})$, the relations (\ref{lukrep1.7}) can be smeared out with such a 
		two-particle test function as follows
\begin{eqnarray}\label{lukrep3.8}
	(a_\kappa \circ a_\kappa) 
	[f^{(2)}] & = &
	\int d^3\vec{p} \int d^3 \vec{q} \, f^{(2)} (\vec{p}, \vec{q})
	(a_\kappa(\vec{p})\circ a_\kappa (\vec{q}))
	\\
	&=&
	\int d^3\vec{p} \int d^3 \vec{q} \, f^{(2)} (\vec{p}, \vec{q})
	a_\kappa(\vec{p}\, e^{\frac{q_0}{\kappa}}, p_0)
	a_\kappa(
	\vec{q} \, e^{-\frac{p_0}{\kappa}},q_0)\Big|_{{p_0=\omega_\kappa (\vec{p})}
	\atop
	{q_0=\omega_\kappa (\vec{q})}}
	\nonumber 
\end{eqnarray}
i.e. firstly we should $\kappa$-multiply the oscillators, and then to integrate.
In such a way we obtain the $\kappa$-deformed 2-particle wave packet describing a cluster which however can not be separated into the product of 
$\kappa$-deformed  one-particle wave packets.

In particular in the formula (\ref{lukrep3.8}) one can perform the change of variables
\begin{equation}\label{lukrep3.9}
	p_i = p_i(\vec{\cal P}, \vec{\cal Q}) \qquad \qquad
	q_i = q_i(\vec{\cal Q}, \vec{\cal P})
\end{equation}
and rewrite (\ref{lukrep3.8}) using new three-momentum variables $\vec{\cal{P}}, \vec{\cal{Q}}$, with the two-particle wave packet function transformed as follows
\begin{equation}\label{lukrep3.10}
	\widetilde{f}^{(2)}(\vec{\cal{P}}, \vec{\cal{Q}}) =
	 {\cal J}
	 \Big[{{p_i,q_i}\atop {P_i,Q_i}}\Big]
	 (\vec{\cal{P}}, \vec{\cal{Q}})\,
	 {f}^{(2)}
	 (\vec{p} (\vec{\cal{P}}, \vec{\cal{Q}}), \vec{q} (\vec{\cal{P}}, \vec{\cal{Q}}))
	 \end{equation}
	 where ${\cal J}$ describes the Jacobian of the transformation (\ref{lukrep3.9}). We would like to stress that the transformations 
	 (\ref{lukrep2.8}), (\ref{lukrep2.13}), (\ref{lukrep2.16})
	 linking two ways of $\kappa$-deforming the oscillators algebra 
	  belong to the class given by relations (\ref{lukrep3.9}).
	 
	 The $\kappa$-deformed multiplication (\ref{lukrep1.7})--(\ref{lukrep1.10}) in momentum space can be transformed into the dual operation in space-time, described by $\kappa$-deformed star product. One can introduce the positive frequency $\kappa$-deformed free field operator, satisfying $\kappa$-deformed  KG equation (see e.g. \cite{lukrep2}--\cite{lukrep3})
\begin{equation}\label{lukrep3.11}
	\varphi^{(+)}_\kappa(x) = \frac{1}{(2\pi)^{3/2}}
	\int \frac{d^3 \, \vec{p}}{\Omega_\kappa(\vec{p})}\,
	a_\kappa(\vec{p}, \omega_\kappa(\vec{p}))
	\, e^{i(\vec{p} \vec{x} - \omega_\kappa(\vec{p})t)}
\end{equation}
where 
\begin{equation}\label{lukrep3.12x}
	\Omega_\kappa = 2 \kappa \sinh \,
	\frac{\omega_\kappa(\vec{p})}{\kappa}
	\quad
	\omega_\kappa (\vec{p}) = 2 \kappa\ \hbox{arc}\sinh \left(
	\frac{\sqrt{\vec{p}^2+M^2}}{2\kappa}
	\right)\, .
\end{equation}
 Let us define
the $\kappa$-deformed star product $\star _\kappa$ 
of the space part of the exponential functions occurring in (\ref{lukrep3.11})
\begin{equation}\label{lukrep3.12}
	e^{ipx}\star_\kappa e^{iqy}_{\ \ \ |x_0=y_0=0}\equiv e^{i \vec{p} \vec{x}} \star _\kappa e^{i \vec{q} \vec{y}}
	= e^{i \left(
		e^{\frac{\omega_\kappa(\vec{q})}{2\kappa}}\vec{p} \vec{x}
		+
	e^{-\frac{\omega_\kappa(\vec{p})}{2\kappa}}\vec{q} \vec{y}
	\right)}\, .
\end{equation}
The field product $\varphi^{(+)} (x) \star _\kappa \varphi^{(-)}(y)$ is well defined only if it is understood as follows
\begin{eqnarray}\label{lukrep3.13}
\varphi^{(+)}_\kappa (x) \star _\kappa \varphi^{(+)}_\kappa(y)
\ \mathop{=}\limits^{ \hbox{\footnotesize def} } \
\frac{1}{(2\pi)^3} \int \frac{d^3 \vec{p}}{\Omega_\kappa(\vec{p})}
\,
\frac{d^3 \vec{q}}{\Omega_\kappa(\vec{q})}
\cr
\cdot a_\kappa(\vec{p},p_0) a_\kappa(\vec{q}, q_0)
\left(
e^{i\vec{p}\vec{x}} \star _\kappa
e^{i \vec{q}\vec{y}}
\right)
e^{i(p_0 x_0 + q_0 y_0)}\Big|_{{p_0=\omega_\kappa(\vec{p})}
\atop
{q_0=\omega_\kappa(\vec{q})}}
\end{eqnarray}
i.e. firstly we perform the $\star_\kappa$-product of exponentials, and then we integrate. We see therefore that
 the notation using the factorized $\star _\kappa$-product structure of fields on lhs of (\ref{lukrep3.13}) is not well justified. More correctly the star product $\star _\kappa$ describes  a $\kappa$-deformed bilocal field

\begin{equation}\label{lukrep3.14}
	\varphi^{(+)}_\kappa (x) \star _\kappa \varphi^{(+)}_\kappa (y) \equiv \varphi^{(+,+)}_\kappa (x;y)\, .
\end{equation}
Let us introduce the relations
\begin{eqnarray}\label{lukrep3.15}
p_i\left(
e^{i \vec{p} \vec{x}} \star _\kappa e^{i \vec{q} \vec{y}}
\right)
= e^{i \frac{\partial^y_0}{2\kappa}} \, \partial^x_i
\left(
e^{i\vec{p}\vec{x}} \star e^{i\vec{q}\vec{y}}
\right)
\cr
q_i\left(
e^{i \vec{p} \vec{x}} \star e^{i \vec{q} \vec{y}}
\right)
= e^{- i \frac{\partial^x_0}{2\kappa}} \, \partial^y_i
\left(
e^{i\vec{p}\vec{x}} \star e^{i\vec{q}\vec{y}}
\right)
\end{eqnarray}
where $i \partial^y_0 = \omega_\kappa(-\Delta^y)$,
$i \partial^x_0 = \omega_\kappa(-\Delta^x)$
 ($\Delta^x = \partial^x_i \partial^x_i$ etc.).
  It is easy to check that the bilocal field (\ref{lukrep3.14}) satisfies the bilocal field equation

\begin{eqnarray}\label{lukrep3.16}
\left[
\Delta^x \, e^{i\frac{\partial^y_0}{\kappa}}- \left(
2\kappa \sinh \frac{\partial^x_0}{2\kappa}\right)^2
\right]
\left[
\Delta^y \, e^{-i\frac{\partial^x_0}{\kappa}} - \left(
2\kappa \sinh \frac{\partial^y_0}{2\kappa}\right)^2
\right]
\varphi_\kappa^{(+,+)} (x;y)=0
\cr
\end{eqnarray}
We see that the $\star _\kappa$-product of two free fields introduces the nonlocal coupling, because 
\begin{equation}\label{lukrep3.17}
	e^{-i\frac{\partial^x_0}{\kappa}}\varphi^{(+,+)}_\kappa
	(\vec{x},x_0; \vec{y},y_0) =
	\varphi^{(+,+)}_\kappa(\vec{x},x_0 +\frac{1}{\kappa}; \vec{y},y_0)
\end{equation}
and the differential operator in Eq.~(\ref{lukrep3.16}) is not factorizable. This non-factorizability reflects the geometric interaction introduced by noncommutative structure of $\kappa$-deformed space-time.

The relation between the $\circ$-multiplication of the field oscillators (see (\ref{lukrep1.7})) and the $\star _\kappa$-multiplication of fields was studied in \cite{lukrep2}. We obtain the relation
\begin{equation}\label{lukrep3.18}
	\varphi^{(+)}(x) \hat{\star }_\kappa
	\varphi^{(+)}(y) =
	\varphi^{(+)}(x) \circ_{\mathrm rel} \varphi^{(+)}(y)
\end{equation}
if we modify in the formula (\ref{lukrep3.13}) the energy-momentum relations 
in accordance with (\ref{lukrep2.9}), i.e. 
\begin{equation}\label{lukrep3.19}
	\begin{array}{lcl}
	\begin{array}{l}
	p_0=\omega_\kappa(\vec{p})
	\cr
		q_0=\omega_\kappa(\vec{q})
	\end{array}
	\quad & \Rightarrow \quad 
	&
		\begin{array}{l}
	p_0=\omega_\kappa(\vec{p} \, e^{-\frac{q_0}{2\kappa}})
	\cr
		q_0=\omega_\kappa(\vec{q}\,  e^{\frac{p_0}{2\kappa}})
	\end{array}
	\end{array}
\end{equation}
and  we understand the multiplication of space-time fields 
by firstly performing  the multiplication of oscillators  and then integrating over the momenta

\begin{eqnarray}\label{lukrep3.20}
\varphi^{(+)}_\kappa (x) \circ \varphi^{(+)}_\kappa(y)
\ \mathop{=}\limits^{ \hbox{\footnotesize def} } \
\frac{1}{(2\pi)^3} \int \frac{d^3 \vec{p}}{\Omega_\kappa(\vec{p})}
\,
\frac{d^3 \vec{q}}{\Omega_\kappa(\vec{q})}
\cr
\cdot\left(
 a_\kappa(\vec{p},p_0) \circ a_\kappa(\vec{q}, q_0)\right)
e^{i(px+ qy)}
\Big|_{{p_0=\omega_\kappa(\vec{p})}
\atop
{q_0=\omega_\kappa(\vec{q})}} 
\end{eqnarray}
The basic relation (\ref{lukrep3.18}) linking two different pictures of $\kappa$-deformation is valid provided that we introduce into the multiplication (\ref{lukrep1.7})  the relativistic numerical factor (see \cite{lukrep2})
\begin{equation}\label{lukrep3.21}
	a_\kappa({p}) \circ_{\text{rel}} a_\kappa({q})
	= e^{\frac{3}{2\kappa}
	(\omega_\kappa(\vec{p}) -\omega_\kappa(\vec{q}))}
	a_\kappa({p}) \circ a_\kappa({q})\, .
\end{equation}

\section{$\kappa$-deformed $n$-particle Cluster and some  Properties of $\kappa$-deformed \\ Oscillators Algebra}

\setcounter{equation}{0}

In  Sect.~3 we described the binary $\kappa$-deformed products of the field oscillators and space-time fields. In order to describe all sectors of $\kappa$-deformed Fock space (\ref{lukrep3.3}) one should introduce the multiplication rules between arbitrary number of oscillator monomials forming the linear basis of algebra ${\cal A}$. First generalization is provided by the extension of (\ref{lukrep1.7}) to the pair of arbitrary monomials of creation oscillators as follows

\begin{eqnarray}\label{lukrep4.1}
 &&
 (a_\kappa({p}_1) \ldots a_\kappa({p}_n))
\mathop{\circ}
 (a_\kappa(q_1) \ldots a_\kappa(q_m)) = 
 \cr 
 &&\qquad \qquad \qquad = \prod\limits^{n}_ {i=1}
 a_\kappa\left(
 \vec{p}_i \,
 e^{\frac{1}{2\kappa} \sum\limits^{m}_{k=1} \omega_\kappa(\vec{q}_k)}
 , p^0_i
 \right)
 \, \prod^{m}_{j=1}
  a_\kappa\left(
  \vec{q}_j \,
 e^{-\frac{1}{2\kappa} \sum\limits^{n}_{l=1} \omega_\kappa(\vec{p}_k)}
 , q^0_j
 \right)
\end{eqnarray}
where $p^0_i=\omega_\kappa(\vec{p}_i), q^0_j=\omega_\kappa(\vec{q}_j)$.
In particular using (\ref{lukrep1.7}) and (\ref{lukrep4.1}) we obtain that
\begin{eqnarray}\label{lukrep4.2}
a_\kappa({p}) \circ 
\left(
a_\kappa ({q}) \circ a({r})
\right)
 \equiv
  a_\kappa ({p})
   \circ
\left(
a_\kappa\left(\vec{q} \, e^{\frac{\omega_\kappa (\vec{r})}{2\kappa}}, q_0
\right)
a_\kappa
\left(
\vec{r} \, e^{-\frac{\omega_\kappa (\vec{q})}{2\kappa}}, r_0\right)
\right)=
\cr
a_\kappa
\left(
\vec{p} \, e^{\frac{1}{2\kappa}(\omega_\kappa (\vec{q})
+ \omega_\kappa(\vec{r}))} , p_0
\right)
\cdot
a_\kappa
\left(
\vec{q} \, e^{\frac{1}{2\kappa}(-\omega_\kappa (\vec{p})
+ \omega_\kappa(\vec{r}))}, q_0
\right)
\cdot
a_\kappa
\left(
\vec{r} \, e^{-\frac{1}{2\kappa}(\omega_\kappa (\vec{p})
+ \omega_\kappa(\vec{q}))} , r_0
\right)
\end{eqnarray}
It is easy to show from (\ref{lukrep4.1}) that the associativity relation
\begin{equation}\label{lukrep4.3}
	a_\kappa({p}) \circ (a_\kappa({q}) \circ a_\kappa({r})) =
	(a_\kappa({p}) \circ a_\kappa({q})) \circ a_\kappa(r) 
\end{equation}
is valid, i.e. one can describe the product (\ref{lukrep4.3}) without marking the binary brackets.

The $\kappa$-multiplication rule (\ref{lukrep4.1}) has the following feature: \
the expressions containing both standard and $\kappa$-deformed multiplications are not associative. In particular from (\ref{lukrep4.1}) follows that for $i\neq k$ or $j \neq l$~\footnote{The relation (\protect\ref{lukrep4.1}) describes a special case of lhs of Eq.~(\protect\ref{lukrep4.4}), when $l=1, j=m$.}

\begin{eqnarray}\label{lukrep4.4}
a_\kappa({p}_1) \ldots a_\kappa({p}_{l-1})
(a_\kappa({p}_l) \ldots a_\kappa({p}_n) \circ a_\kappa({q}_1) \ldots 
a_\kappa({q}_j))
a_\kappa({q}_{j+1}) \ldots a_\kappa({q}_m) \qquad
\cr
\quad \neq
a_\kappa({p}_1) \ldots a_\kappa({p}_{k-1})
(a_\kappa({p}_k) \ldots a_\kappa({p}_n) \circ a_\kappa({q}_1) \ldots 
a_\kappa({q}_l))
a_\kappa({q}_{l+1}) \ldots a_\kappa({q}_m) \, .
\end{eqnarray}
Such a property leads to peculiar factorization properties of the binary oscillator relations in the multiparticle sectors ${\cal H}^\kappa_n$ with $n>2$ generated from the vacuum by $n$-fold $\circ$-products of the creation oscillators:
\begin{equation}
\mathcal{H}_n^\kappa:~~|\vec{p}_1>\otimes_\kappa\dots\otimes_\kappa|\vec{p}_n>\equiv a(p_1)\circ\dots\circ a(p_n)|0>\,.
\end{equation}

If $n=3$ from  relation (\ref{lukrep1.7}) by standard multiplication  one gets  the identity
\begin{equation}\label{lukrep4.5}
	a_\kappa(r_0, \vec{r}) \cdot (a_\kappa({p}) \circ a_\kappa({q}) 
	-a_\kappa({q}) \circ a_\kappa({p})) =0 \, .
\end{equation}
However, if we use the $\kappa$-multiplication, we obtain
\begin{equation}\label{lukrep4.6}
	a_\kappa({r}) \circ (a_\kappa({p}) \circ a_\kappa({q})
	- a_\kappa({q}) \circ a_\kappa({p})) = 0 \, .
\end{equation}
The relations (\ref{lukrep4.6}) is not equivalent to the relations (\ref{lukrep4.5}).
 The explicit form of the relation (\ref{lukrep4.6}) leads on the basis of (\ref{lukrep4.2}) to the formula
\begin{eqnarray}\label{lukrep4.7}
a_\kappa(\vec{r}\, e^{\frac{p_0+q_0}{2\kappa}}, r_0)
(
a_\kappa( \vec{p} \, 
e^{\frac{- r_0+q_0}{2\kappa}}, p_0
)
a_\kappa ( \vec{q}\,
e^{\frac{- r_0 - p_0}{2\kappa}}, 
q_0
)
\cr
\qquad \qquad \qquad - \
a_\kappa(\vec{q} \,
 e^{\frac{- r_0+q_0}{2\kappa}}, q_0
)
a_\kappa
( \vec{p} \,
e^{\frac{- r_0- p_0}{2\kappa}}, 
p_0
))
=0\, .
\end{eqnarray}

We see from (\ref{lukrep4.7}) that in three-particle sector ${\cal H}^\kappa_3$ the relation (\ref{lukrep1.7}) takes the modified form
\begin{equation}\label{lukrep4.8}
	a_\kappa({p} (r_0)) \circ a_\kappa({q} (r_0)) 
	\doteq 
	a_\kappa({q}(r_0)) \circ a_\kappa({p} (r_0))
\end{equation}
where 
\begin{equation}\label{lukrep4.9}
	\vec{p} (r_0) = \vec{p} \, e^{-\frac{r_0}{2\kappa}}
	\qquad \qquad
		\vec{q} (r_0) = \vec{q} \, e^{-\frac{r_0}{2\kappa}}
\end{equation}
describes the three-momenta shifted by the factor depending on the energy $r_0=\omega_\kappa(\vec{r})$ of the third particle.

Interestingly enough, if we factorize the relation (\ref{lukrep1.7}) in ${\cal H}^\kappa_3$ in different way.
\begin{equation}\label{lukrep4.10}
	\left(
	a_\kappa({p}) \circ a_\kappa ({q}) - a_\kappa ({q}) \circ a_\kappa({p})
	\right) \circ a_\kappa({r}) =0
\end{equation}
the factorized form of binary relations takes different form, with opposite sign of the energy of third particle
\begin{equation}\label{lukrep4.11}
	\left(
	a_\kappa({p} (-r_0)) \circ \, a_\kappa ({q} (-r_0)) =
	 a_\kappa ({q} (-r_0)) \circ a_\kappa({p} (-r_0)\right)\, .
\end{equation}
The explanation of such modifications of our binary  oscillator relations follow from the fact that the $\kappa$-multiplication prescription is not universal, i.e. it is different in two sectors ${\cal H}^\kappa_n$, ${\cal H}^\kappa_n$ if $n\neq m$ . Originally we derived the oscillator binary relations (\ref{lukrep1.7})--(\ref{lukrep1.15}) in ${\cal H}^\kappa_2$; if 
 we factorize the binary relations  in ${\cal H}^\kappa_3$, we arrive at the relations (\ref{lukrep4.6}) or (\ref{lukrep4.10}).
The factorized binary relation in ${\cal H}^\kappa_3$ expressed by the multiplication rules from ${\cal H}^\kappa_2$ (see (\ref{lukrep1.7})--(\ref{lukrep1.15})), take the  forms (\ref{lukrep4.8}) or (\ref{lukrep4.11}). Using physical arguments, if we assume that the $\kappa$-multiplication describes some geometric interaction between particles, one can argue that if we change the number of particles and pass from the sector ${\cal H}^\kappa_n$ to ${\cal H}^\kappa_m$ 
 $(n\neq m)$, we  modify as well the interactions what leads to the modification of the $\kappa$-deformed multiplication rule.

The $n$-fold $\kappa$-deformed multiplication rule for the $\kappa$-deformed oscillators
\begin{equation}\label{lukrep4.12}
m_\kappa (A_1 \otimes \ldots \otimes A_n) = A_1 \circ \ldots \circ A_n
\end{equation}
 is different therefore in every sector ${\cal H}^\kappa_{n}$.
  We extend the formula (\ref{lukrep4.2}) to any  $\kappa$-deformed   product  of single oscillators as follows\footnote{For simplicity  of our considerations
 in this Section we deal only with the subalgebra 
  of creation operators. Inclusion of annihilation operators does not  produce new difficulties.}
\begin{eqnarray}\label{lukrep4.13}
&&
a_\kappa({p}_1) \circ \ldots \circ a_\kappa({p}_n)=
a_\kappa
\left( \vec{p}_1 \, e^{\frac{1}{2\kappa}\sum\limits^{n}_{j=2} p^j_0}
, p^1_0
\right)\ldots
\\
&&
\quad \quad
\ldots 
a_\kappa
\left( \vec{p}_k \, e^{\frac{1}{2\kappa}(-\sum\limits^{k-1}_{l=1} p^l_0
 + \sum\limits^{n}_{j=k+1} p^j_0)}
, 
p^k_0 \right) \ldots
a_\kappa
\left( \vec{p}_n \, e^{-\frac{1}{2\kappa}\sum\limits^{n-1}_{l=1} p^l_0}
, 
p^n_0 \right)\,.
\nonumber
\end{eqnarray}
In order to prove the associativity relation one should  extend  the formula (\ref{lukrep4.13}) to the $\circ$-products of arbitrary monomials of the oscillators, i.e. one should generalize the formula (\ref{lukrep4.1}) to any number of $\circ$-multiplications. Such formula is easy to write down if we observe the rule that any oscillator $a_\kappa(\vec{p}, p_0)$ located in the product to the left of a given oscillator $a_\kappa(\vec{q}, q_0)$ shifts its threemomentum $\vec{q}$ by multiplicative factor $e^{-\frac{p_0}{2\kappa}}$;  the location of $a_\kappa(\vec{p}, p_0)$  on the right side multiplies $\vec{q}$ by the factor $e^{\frac{p_0}{2\kappa}}$.

Finally we shall stress that the associative $\kappa$-deformed multiproducts of oscillators is equivalent in space-time to suitable associative extension of $\hat{\star }_\kappa$-star product. The relation (\ref{lukrep3.18}) can be generalized as follows:
\begin{equation}\label{lukrep4.14}
	\varphi^{(+)}_\kappa(x_1) 
	{\hat{\star }}_\kappa
	\ldots
	{\hat{\star }}_\kappa
	\varphi^{(+)}_\kappa(x_n) =
	\varphi^{(+)}_\kappa(x_1) \circ \ldots \varphi^{(+)}_\kappa(x_n) \,,
\end{equation}
where the multiplications of oscillators and star products of exponential functions should be performed  in (\ref{lukrep4.14}) before the momenta integrations. 

We add the following comments in relation with the formula (\ref{lukrep4.14}):

i) The unusual properties of the multiplication is reflected as well in the properties of the ${{\star }}_\kappa$-product for $n$ fields. The
definition of ${\hat{\star }}_\kappa$ depends on all the  fields in the product (\ref{lukrep4.14}), what reflects the properties of original noncommutative fields $\varphi_\kappa(\hat{x}_1)\ldots \varphi_\kappa(\hat{x}_n)$ which depend on  noncommutative space-time coordinates satisfying the relations\footnote{For the canonical noncommutativity ($\theta_{\mu\nu} = \hbox{const}$) see for comparison the $n$-fold star products in \cite{lukrep11}.}
\begin{equation}\label{lukrep4.15}
	\left[
	\hat{x}^\mu_i, \hat{x}^\nu_j
	\right]\neq 0
\end{equation}
for any pair of $i,j$. If e.g. for $i=1$ we have noncommutativity (\ref{lukrep4.15}) for $j=2,3 \ldots n$, this algebraic property reflects the presence of the geometric interaction of the first field with all remaining $n-1$ fields.

ii) The $\kappa$-deformed oscillator multiplication  through the dependence of $i$-th  particle  on the energies of remaining \hbox{$n-1$} particles introduces the  inseparable clusters of $n$-particles. Such a property implies that the $\star _\kappa$-products  and $\circ$-products  (see (\ref{lukrep4.14})) describe in fact the $n$-local field $\varphi^{(+, \ldots +)}_\kappa(x_1 \ldots x_n)$,  which satisfies the nonfactorizable  field equation in the $n$-fold product of Minkowski spaces  generalizing the \, Eq.~(\ref{lukrep3.16}).

\section{Final Remark}

\setcounter{equation}{0}

The introduction of noncommutative coordinates  leads to the geometric interactions between the noncommutative fields. Even in the case of the simplest canonical noncommutativity (see e.g. \cite{lukrep12})
\begin{equation}\label{lukrep5.1}
\big[ 
\hat{x}_\mu , \hat{x}_\nu \big] = i \frac{\theta_{\mu\nu}}{\kappa^2}
\qquad 
\theta_{\mu\nu} = \hbox{const}
	\end{equation}
	the Moyal-Weyl star product of fields

\begin{equation}\label{lukrep5.2}
	\varphi_0 (x) \star _\theta \varphi_0 (y)
	= \varphi_0 (x) \, \exp\Big\{i 
	\frac{\vveecc{\partial}_x^\mu \, \theta_{\mu\nu}
 \vec{\partial}^\nu_y}{\kappa^2}
	\Big\} \varphi_0(y)
\end{equation}
	is not factorizable, i.e. if we define
	
\begin{equation}\label{lukrep5.3}
	\varphi_0(x)  \star _\theta \varphi_0 (y) =
	\int d^4p \int d^4q \ \phi (p,q) \,e^{ipx+qy}
\end{equation}
we obtain that $\phi(p,q) \neq \phi_1(p)\cdot \phi_2(q)$. 

In the case of
 deformation (\ref{lukrep5.3})  the $\star _\theta$-product
of free fields satisfies the ``free'' bilocal equation
\begin{equation}\label{lukrep5.4}
	\left(
	\square_x - m^2
	\right)
	\left(
	\square_y - m^2
	\right)
	\left(
	\varphi_0(x) \star _\theta \varphi_0(y)
	\right) = 0 \ .
	\end{equation}
	If we proceed to the  $\kappa$-deformation, the noncommutativity of $\kappa$-Minkowski space leads to the modification of KG operators

\begin{equation}\label{lukrep5.5}
	\square_x   \mathop{\longrightarrow}\limits_{ \kappa < \infty }
	\Delta_x  - \left(
	2\kappa \, \sin \, \frac{\partial^x_0}{2\kappa}
	\right)^2
\end{equation}
as well as the appearance of the ``cross-terms'' in the bilocal differential ope\-ra\-tor (see (\ref{lukrep3.16})).

Concluding, we recall that in this lecture we presented
 the following two new features:

i) In Sect.~2 we did show the equivalence between two ways of describing  $\kappa$-deformed binary oscillator relations:  by means of $\kappa$-deformed multiplication and by the use of $\kappa$-deformed flip operator.

ii) In Sect.~3 and 4 we described some physically relevant properties of $\kappa$-deformed  algebra of space-time free fields as well as $\kappa$-deformed  field  oscillator algebra. Both deformations are related (see (\ref{lukrep3.18}) and (\ref{lukrep4.13})) and they lead to nonfactorizable (entangled) $\kappa$-deformed products of fields and nonfactorizable (entangled) $\kappa$-deformed $n$-particle states. This lack of  factorizability reflects the presence of geometric interaction between particle states which is algebraically described  by the $\kappa$-deformation.

\subsection*{Acknowledgments}
The author would like to thank Mariusz Woronowicz for discussions. This paper has been supported by Polish Ministry of Science and Higher Education grant NN 202318534.

\end{document}